    \documentclass[prl,aps,twocolumn,showpacs]{revtex4}
\usepackage[dvips]{graphicx}
\usepackage{dcolumn}
\newcommand{\be}{\begin{equation}}
\newcommand{\ee}{\end{equation}}
\newcommand{\bea}{\begin{eqnarray}}
\newcommand{\eea}{\end{eqnarray}}
\newcommand{\nn}{ \nonumber}

\begin{document}
\topmargin=-20mm

\title{Low Temperature Electronic Transport through Macromolecules and Characteristics of Intramolecular Electron Transfer}

\author{ Natalya A. Zimbovskaya}

\affiliation{ 
Department of Physics and Electronics, University of Puerto Rico - Humacao, PR 00791 } 

 \begin{abstract}
Long distance electron transfer (ET) plays an
important part in many biological processes.  Also, fundamental understanding of ET processes  could give grounds for designing miniaturized electronic devices. So far experimental data on the ET mostly concern ET rates which characterizes ET processes in whole. Here, we develop a different approach which could provide more information about intrinsic characteristics of the long-range intramolecular ET. A starting point of the studies is an obvious resemblance between ET processes and electric transport through molecular wires placed between metallic contacts. Accordingly, the theory of electronic transport through molecular wires is applied to analyze characteristics of a long-range electron transfer  through molecular bridges. Assuming a coherent electron tunneling to be a predominant mechanism of ET at low temperatures, it is shown that low-temperature current-voltage characteristics could exhibit a special structure, and the latter contains information concerning intrinsic features  of the intramolecular ET.  Using the Buttiker dephasing model within the scattering matrix formalism we analyze the effect of dephasing on the electron transmission function and current-voltage curves.
 \end{abstract}

\date{\today} 
\maketitle


\section{I. Introduction and Background}
\vspace{0mm}

It is well known that long distance electron transfer (ET) plays an important part in many biological processes such as signal transduction across membranes, photosynthesis, enzyme cathalysis and other reactions in
biological cells \cite{one,two}. A fundamental understanding of these reactions also has substantial implications for further miniaturization of electronic
devices \cite{three,four,five}. Theoretical and experimental investigation of ET processes in biological macromolecules such as proteins and DNA lasts more than
three decades but it still remains a very active area of research. ET kinetic data on macromolecules are increasingly available (for recent experimental results, see e.g. \cite{six,seven} and references therein).

It has been established that molecular ET is essentially a combination of nuclear environment fluctuations and electron tunneling. Due to the large distances between donor and acceptor subsystems in macromolecules $(\sim 1 nm), $ a direct electronic coupling between them is extremely small, so the ET is provided by  intervening molecular bridges, giving rise to  sets of intermediate states for the electron tunneling. The expression for the ET rate including both electronic and
nuclear factors was first proposed by Marcus \cite{eight,nine,ten} and  can be written as follows:
     \begin{equation} 
K_{ET} = K_{el} K_n \nu_n \ . 
    \label{e1}       \end{equation}
 Here, $ K_{el}$ is the electron transmission coefficient, $ K_n $ is the nuclear transition or Franck-Condon factor, and $ \nu_n $ is the effective nuclear vibration frequency.

The enormous size of biological molecules makes straightforward calculations of $ K_{el}$ extremely difficult. As a result of numerous theoretical studies, some methods have been developed to simplify the
analysis of ET processes. These methods are mostly based on a general model of primary or highly probable pathways (see e.g. Refs. \cite{eleven,twelve,thirteen,fourteen,fifteen,sixteen,seventeen}).  The point of this approach is that an electron involved in a long range ET process, with a high probability follows a few primary tunneling pathways moving through the macromolecule. Therefore large portions of the original macromolecule do
not significantly contribute to the electron transfer, and can be excluded from further calculations.   Keeping only these primary pathways, one can reduce original molecular bridges to much simpler chain-like structures.
This enables to develop a numerical analysis of elecron transport along the chains.

  Assuming the coherent tunneling (superexchange mechanism) to predominate  one can treat a long-range ET process as a sequence of tunnelings between orbitals of the considered donor-bridge-acceptor system. The latter are sites of the system, and any atom included into consideration is represented as a set of sites
corresponding to its states associated with valence electrons. The electron transmission coefficient $ k_{el} $ equals $ |T_{DA}|^2 $, where $T_{DA}$ is the transition amplitude between donor and acceptor
subsystems. The transition amplitude can be presented in the form:
     \be 
T_{DA}^{mn} = T_{mn}^{0} + \sum_{ij} {\cal D}_{mi} G_{ij} {\cal A}_{jn}.
           \ee  
 Here, $T_{mn}^{0}$ is the direct coupling of the donor site "m" to the acceptor site "n" which is usually weak in macromolecules and therefore can be neglected. The summation in the second term  is carried out over the bridge sites involved in the ET process; $ {\cal D}_{mi} $ and $ {\cal A}_{jn}$ are the couplings of
the donor/acceptor to the bridge; $ G_{ij} $ are matrix elements of the Green's function between bridge sites $ i$ and $ j $:
     \be 
G_{ij} = <i |(E - H)^{-1}|j>
           \ee  
 where, $ H $ is the Hamiltonian of the bridge, and $ E $ is the tunneling energy.

This method of calculation of $T_{DA} $ was repeatedly employed in numerous papers  \cite{eighteen}. Obtained results were succesfully used to justify semiempirical expressions for $k_{el}$ which were applied for interpretation of experimental results, such as the pathway model (see Ref. \cite{six}), or the average packing density model  \cite{nineteen}.

Thus far, insufficient attention was paid to systematically analyze the contribution of donor/acceptor coupling to the bridge to the long-range ET
processes. Usually, in theoretical studies concerning ET processes both donor and acceptor are presented as single sites, whereas in realistic biological macromolecules the donor/acceptor subsystem often has a complex structure and includes a set of sites providing effective coupling to the bridge. Correspondingly, the bridge has a set of entrances and a set of exits which an electron involved in the ET process can employ. At
different values of the tunnel energy $ E$ different sites of donor and acceptor subsystems can give predominant contributions to the ET. In other words, at different values of $ E $ preferred entrances to the bridge and exits from it, as well as primary pathways for electrons, differ.

 ET processes in complex molecular systems occur in a highly disordered nuclear/solvation enviroment. Stochastic fluctuations of the environment electric potential cause the electronic phase-breaking effect. Also, incoherent component in the electron transport could originate from the electron scattering on intramolecular vibrational modes \cite{twentythree} and due to some other dissipative processes 
\cite{twentyfour}. The dephasing could strongly affect the character of the ET processes destroying electron pathways in the molecules and providing a transition from coherent superexchange to the completely incoherent sequental hopping mechanism of the electron transfer \cite{twentyfive,twentysix,twentyseven,twentyeight}.
The effect of inelastic scattering in the ET processes is often described within the Buttiker approach \cite{twentynine}. The electron propagation along the molecular bridge is considered as a multichannel scattering problem which is treated using the scattering matrix formalism \cite{twentyseven,thirty,thirtyone,thirtytwo,thirtythree,thirtyfour}. Also, some other formalisms are employed to treat the inelastic scattering effects on the ET characteristics which are discussed elsewhere \cite{thirtyfive}.

 ET processes reveal a strong resemblance with the electron conductance through molecular wires \cite{thirtysix}. This resemblance is a starting point for the present work. We believe that studies of electronic transport through macromolecules can provide important information concerning quantum dynamics of electrons participating in the ET processes. In this paper a simple
approach is proposed which shows that some intrinsic characteristics of the intramolecular ET such as pathways of tunneling electrons and distinctive features of donor/acceptor coupling to the bridge at different
values of tunnel energy can be obtained in experiments on low-temperature electrical transport through the corresponding molecules. These data may be available under moderate electronic dephasing \cite{twentyseven}, when the structure of electron transmission function containing the desired information is not completely
washed out. A similar approach could also be developed to analyze characteristics of intermolecular ET processes in extended bio-systems.

 The organization of this paper is as follows. In Sec. II the  adopted model is introduced to analyze coherent electron tunneling through a molecular bridge. Within this model we take into account complex structure of donor and acceptor subsystems which is typical for realistic macromolecules. 
In the Sec. III we analyze how electronic scattering influences the transmission function. We follow Buttiker dephasing model, and we apply it to a simple single-site bridge. This enables to solve the problem mostly analytically. In the last Section  we discuss how these results could be used to get additional information concerning intrinsic characteristics of ET processes. 

\section{II. The coherent electron transport}

Existing theories of electrical transport through macromolecules are mostly based on a very simple model simulating both electrodes by semi-infinite tight-binding chains attached to the ends of the molecule. The latter
is also simulated as a single tight-binding chain of sites (see e.g. \cite{twentyseven,thirty,thirtyone,thirtysix,thirtyseven,thirtyeight,thirtynine}). Here we adopt a resembling model to analyze the electric transport through realistic macromolecules including donor, bridge, and acceptor subsystems, but we
take into account that donor and acceptor subsystems are complex and include many sites which exhibit a significant coupling to the bridge.

To simplify first steps of further calculations we separate out the effect of electron dephasing, assuming coherent electron tunneling through the macromolecule
to be the predominant mechanism of electrical transport. Correspondingly, we treat the electron transport through the molecule as a sequence of tunnelings between potential wells. Each well represents one atomic
orbital of the donor-bridge-acceptor system. An electron which participates in the transport is treated as moving in an array of these potential wells which
 are the sites of the array.  Within this approach,
any atom is represented by a set of sites corresponding to its states and we do not distinguish between sites corresponding to the same atom and sites corresponding  to different atoms of the donor-bridge-acceptor
system. Sometimes it is more convenient to treat the bridge as consisting of atomic groups such as, for example, the methyl groups CH$_2$ instead of
single atoms. Then any such atomic group can be represented as a set of sites, each corresponding to a molecular orbital for the group. For further simplification we assume that intraatomic hopping integrals are smaller that interatomic ones. This enables us to consider electron transfer between different sites as its tunneling between different atoms.

Assume that sites $"i"$ are associated with those atoms of the bridge which have the maximum coupling to the donor, and sites $"j"$ are associated with atoms strongly interacting with the acceptor. Neglecting
the intraatomic hopping, we can treat any site $"i"$ which represents an atom at the entrance of the bridge as the origin of a chain along which an electron moves from donor to acceptor, and any site $"j"$ as the end of the chain. Thus, we have a set of chains (pathways) for an electron moving along the bridge. In further analysis, we concentrate on the case when the chains weakly interact with each other,  so we can consider them separately, as illustrated in the diagram below.

\vspace{14mm}

{\small
\begin{picture}(0,0)(10,0)

\put(0,0){\framebox(36,28)}
\put(70,0){\framebox(36,28)}
\put(140,0){\framebox(36,28)}
\put(205,0){\framebox(36,28)}
\put(2,16){D}
\put(232,16){A}

\put(18,5){\oval(8,4)} 
\put(18,14){\oval(8,4)}
\put(18,23){\oval(8,4)}

\put(30,5){\circle*{3}} \put(22,5){\vector(1,0){66}}
\put(30,14){\circle*{3}} \put(22,14){\vector(1,0){66}}
\put(30,23){\circle*{3}} \put(22,23){\vector(1,0){66}}

\put(89,5){\circle*{3}} \put(89,5){\vector(1,0){26}}
\put(89,14){\circle*{3}}\put(89,14){\vector(1,0){26}}
\put(89,23){\circle*{3}}\put(89,23){\vector(1,0){26}}
\put(117,2){$\cdots$}
\put(117,11){$\cdots$}
\put(117,20){$\cdots$}

\put(158,5){\circle*{3}} \put(131,5){\vector(1,0){26}}
\put(158,14){\circle*{3}}\put(131,14){\vector(1,0){26}}
\put(158,23){\circle*{3}}\put(131,23){\vector(1,0){26}}
\put(158,5){\vector(1,0){55}}
\put(158,14){\vector(1,0){55}}
\put(158,23){\vector(1,0){55}}

\put(214,5){\circle*{3}}\put(214,5){\line(1,0){6}}
\put(214,14){\circle*{3}}\put(214,14){\line(1,0){6}}
\put(214,23){\circle*{3}}\put(214,23){\line(1,0){6}}
\put(224,5){\oval(8,4)}
\put(224,14){\oval(8,4)}
\put(224,23){\oval(8,4)}

\end{picture}}
\vspace{6mm}

We also simulate the donor/acceptor subsystems as sets of semi-infinite tight-binding homogeneous chains. Each chain is attached to a site of donor/acceptor which can be effectively coupled to the bridge. As before, we assume that the chains do not interact. The adopted simple model does not enable us to carry out a proper treatment of coupling of electrodes to the molecule at metal-molecule-metal junctions
which is a nontrivial and intricated problem 
\cite{thirtysix,forty,fortyone,fortytwo}. Here, however, we concentrate on the analysis of electron tunneling from the donor to the acceptor through the bridge. It seems reasonable to conjecture that intrinsic features and characteristics of this process in the molecules
with complex donor and acceptor subsystems do not strongly depend on details of coupling of the electrodes to the donor/acceptor due to a comparatively large size and complicated structure of these subsystems.

In the following calculations we start from a tight-binding Hamiltonian  for a single chain included into the bridge:
  \begin{equation} 
H = H_0 + H_1
   \label{e2}
       \end{equation}
 whose matrix elements between states $<k|$ and $|l>$ corresponding to the $k$-th and $l$-th sites of the chain (the states are supposed to be orthogonal) are given by:
 \begin{equation} 
 (H_0 )_{kl}= \alpha_k\delta_{kl} \  ; \qquad\qquad
(H_1 )_{kl}=  {\cal V}_{kl}\ ,
     \label{e3}
           \end{equation}
 where $\delta_{kl}$ is the Kronecker symbol, ${\cal V}_{kl}$= 0 when $k = l, $ and only states associated with valence electrons are considered.  The diagonal matrix elements $\alpha_k$ are ionization energies of electrons
at sites $k$, and ${\cal V}_{kl} = {\cal V}_{lk}$ includes both direct and exchange energy contributions for an electron to transfer between the $k$-th and $l$-th sites.  As well as it was carried out in \cite{thirtyseven,thirtyeight} we also take into account self-energy corrections arising due to the coupling of the donor $(H_{\cal D})$ and acceptor $(H_{\cal A})$ to the bridge. As a result, we arrive at the effective
Hamiltonian for the chain, i.e.,
  \begin{equation} 
H_{eff}= H + H_{\cal D} + H_{\cal A}\ .
   \label{e4}
       \end{equation}
 Assuming that the chosen chain starts at the $i$-th site and ends at the $j$-th site and generalizing the results of \cite{thirtyone} for the case when we have several donor and acceptor sites coupling to the bridge, we obtain  nonzero matrix elements of $ H_{\cal D}$ and $H_{\cal A}$ in the form
  \begin{equation} 
(H_{\cal D})_{ii} =
(\Sigma_{\cal D} )_{i}=\sum_m \frac{{\cal D}_{mi}^{\
2}}{E-\epsilon_m - \sigma_m}\ ,
      \label{e5}
         \end{equation}
  \begin{equation} 
(H_{\cal A})_{jj} =
(\Sigma_{\cal A})_{j}=\sum_n \frac{{\cal A}_{jn}^{\
2}}{E- \epsilon_n - \sigma_n}\ .
  \label{e6}
        \end{equation}
  Here, ${\cal D} _{mi}$ and ${\cal A}_{jn}$ are respectively coupling strengths between the $m$-th donor site or the $ n $-th acceptor site and the $i$-th or $j$-th site of the bridge, and $\sigma_{m,n} = \frac{1}{2}
\left\{ \theta_{m,n} - i \sqrt{4 \gamma_{m,n}^2 - \theta_{m,n}^2}\right\}$ are  self-energy corrections of the semi-infinite chains attached to the corresponding sites \cite{thirty}.  The parameters $\theta _{m,n} = E - \epsilon_{m,n}$; $\epsilon_{m,n}$ and $\gamma_{m,n}$ are  ionization energies of electrons at the corresponding donor/acceptor sites, and  nearest-neighbor hopping integrals for the chains. Summation in Eqs.\ (7) and (8) is carried out over all donor/acceptor sites coupled to the bridge.  When $E - \epsilon_{m,n}$ is smaller than $\gamma_{m,n}$, we have $\sigma_{m,n} \approx -i \gamma_{m,n}$.  In the opposite limit $ \mid E - \epsilon_{m,n} \mid >> \gamma_{m,n}$ this quantity takes on values close to zero.
          
Due to the presence of the self-energy corrections the eigenvalues of the effective Hamiltonian (6) include imaginary parts which represent broadening of the bridge energy levels $E_i$ originating from the coupling
of the bridge to the donor and acceptor systems. The energy levels are further broadened  if we take into account scattering processes in the bridge.
 
Treating the long-range ET processes, we can neglect the direct coupling of donor sites to acceptor sites.  The probability amplitude for the transition between the $m$-th donor site and the $n$-th acceptor site is given by (2), and $G_{ij}$ is the matrix element of the Green's function corresponding to $H_{eff}$.   An electric tunneling current $I$ flowing from donor to acceptor through the bridge in the presence of a small applied voltage $V$ has the form:   
    \begin{equation} 
 I= \frac{2 \pi e}{\hbar} \sum_{k,l} f \left( E_m \right) \lbrack 1-
f \left( E_n + eV \right)\rbrack \mid T_{mn}^2 \mid \delta \left(E_m - E_n   
\right)
 \label{e9}
          \end{equation}
 where $f(E)$ is the Fermi function, $-e$ is the charge of the electron. Assuming that the applied potential varies linearly in the molecule, we can use the approximation $V = V_0 L_b/L_{mol} $, where $L_b$ and $L_{mol}$ are the lengths of the bridge and the whole molecule, respectively, and $V_0$ is a voltage applied across the whole molecule. Starting from this expression (9), and following a usual way we arrive at the standard formula \cite{thirtysix}: 
   \begin{equation} 
I = \frac{e}{\pi \hbar}\int_{- \infty}^{ \infty}dE\ T(E) \lbrack f \left(
E - {\mu}_1
\right)- f \left( E - {\mu}_2 \right) \rbrack\ ,
 \label{e10}
        \end{equation}
 where the chemical potentials ${\mu}_1$ and ${\mu}_2$ are determined by the equilibrium Fermi energy of the bridge $E_F$ and the effective voltage $V$ across the bridge \cite{thirtyeight}, i.e.
  \[
{\mu}_1 = E_F + \left( 1 - \eta \right) eV;
\qquad  {\mu}_2 = E_F - \eta e V. \]
 Here, the parameter $ \eta $ characterizes how the voltage $ V $ is divided between the two ends of the bridge. The electron transmission function included in the expression (10) is given by the formula
  \begin{equation} 
T(E) = 4 \sum_{i,j} \Delta_i |G_{ij}|^2 \Delta_j \ .
   \label{e11}
      \end{equation}
 
The summation in (11) is carried over states $ <i| $ at the entrances to the bridge and states $ |j> $ corresponding to the exits from the bridge to the acceptor subsystem, therefore contributions from all possible pathways are contained here. The quantities $ \Delta_{i,j} $ are imaginary parts of the self-energy corrections $ \sum_{\cal D,A}, $ namely:
  \begin{equation} 
\Delta_i = \mbox{Im} (\Sigma_{\cal D})_{ii} \ ; \qquad
\Delta_j = \mbox{Im} (\Sigma_{\cal A})_{jj} \ . 
               \label{e12}
      \end{equation}
 
 We see that the dependence of the electron transmission function on energy is determined by the contributions from different donor/acceptor sites, as well as from the Green's function matrix elements corresponding to different chains included into the bridge subsystem. When the tunnel
energy $E$ takes on a value close to $\epsilon_{m,n}$ or to one of the poles of the Green's function, the relevant term in (11) can surpass all remaining terms. Thus, for different values of $E $ different donor/acceptor sites and different pathways can predominate in the electron transfer through the bridge.

This model was applied in numerical simulations of the electron transport in a porphyrin-nitrobenzene molecule \cite{fortythree,fortyfour}. The latter was chosen for it has complex donor and acceptor subsystems connected with a relatively simple bridge. It was shown that low temperature $I-V$ characteristics exhibited a step-like behavior. The steps originate from the structure of the electron transmission $ T (E) $ within the energy range where  $ f(E - \mu_1) - f (E- \mu_2) $ differs from zero. The steps disappear as the temperature rises. Also, the structure of $I-V$ curves  could be significantly affected due to electronic dephasing analyzed below.

\section{III. dephasing effects}

Nuclear motions existing in realistic molecules give rise to the electronic phase-breaking effect. This affects electron transmission function and the electron current through the molecule. When the dephasing is strong (e.g. within the strong thermal coupling limit), the predominant ET mechanism is sequental hopping which replaces the coherent tunneling dominating at weak dephasing. Typically, ET processes occur within an intermediate regime, when both coherent and incoherent contributions to the electron transmission are manifested. The general approach to ET studies in the presence of dissipation is the reduced dynamics density matrix formalism (see e.g. \cite{fortyfive,fortysix}). 
This microscopic computational approach has advantages of being capable to  treat transition from nonadiabatic to adiabatic ET and to provide the detailed dynamics informations. However, these informations are usually much redundant as far as stationary ET processes are concerned. Therefore we choose an alternative approach using the scattering matrix formalism and phenomenological Buttiker dephasing model \cite{twentynine,thirty}. This enables us to analytically treat the problem, and the results agree with those obtained by means of more sophisticated computational methods, as it was demonstrated in the earlier work \cite{twentyseven}.

To proceed analytically, we restrict further consideration with the case of a single-site bridge. We assume that the bridge is coupled to a phase randomizing scatterer. Following the Buttikers work we simulate the scatterer as shown below. Here, we present the ET process by tunneling through two barriers separating the donor and acceptor from the bridge site where the dephasing could occur.  The squares represent tunnel barriers, and the triangle imitates a scatterer coupling the bridge to a dissipative electron reservoir.

This simple model could be succesfully applied to visualize and analyze various phase-breaking processes such as scattering on vibrionic phonon modes in the molecule and/or scattering on the phonons which represent the thermal surrounding. All information concerning characteristics of the scattering processes is contained in the dephasing parameter $ \epsilon $ introduced below

{\small
\vspace{14mm}

\begin{picture}(0,0)(-25,0) 
\put(10,0){\framebox(30,30)}
\put(140,0){\framebox(30,30)}

\put(60,4){\line(1,0){60}} 
\put(60,4){\line(5,4){30}}
\put(120,4){\line(-5,4){30}}
\put(70,4){\line(0,-1){30}}
\put(110,4){\line(0,-1){30}}
\put(66,-4){\vector(0,-1){15}} \put(54,-14){$a_4'$}
\put(74,-19){\vector(0,1){15}} \put(77,-14){$a_4 $}
\put(106,-19){\vector(0,1){15}} \put(94,-14){$a_3$}
\put(114,-4){\vector(0,-1){15}} \put(117,-14){$a_3'$}

\put(10,15){\line(-1,0){25}}  
\put(40,15){\line(1,0){34}}
\put(140,15){\line(-1,0){34}}
\put(170,15){\line(1,0){25}}
\put(60,-26){$ ....................... $}
\put(72,-44){Reservoir}
\put(67,-35){\normalsize 4} \put(108,-35){\normalsize 3}
\put(-21,12){\normalsize 1} \put(197,12){\normalsize 2}

\put(-10,20){\vector(1,0){15}} \put(-5,25){$b_1$}
\put(5,10){\vector(-1,0){15}} \put(-5,0){$b_1'$}
\put(45,20){\vector(1,0){15}}  \put(46,25){$a_1$}
\put(60,10){\vector(-1,0){15}} \put(46,0){$a_1'$}
\put(135,20){\vector(-1,0){15}} \put(125,0){$a_2'$}
\put(120,10){\vector(1,0){15}}  \put(125,25){$a_2$}
\put(190,20){\vector(-1,0){15}} \put(180,25){$b_2$}
\put(175,10){\vector(1,0){15}}  \put(180,0){$b_2'$}

\end{picture}}
\vspace{18mm}

 The scatterer coupled to the bridge is described with the $ 4 \times 4 $ scattering matrix $ s $ relating outgoing wave amplitudes $ a_1', a_2',a_3', a_4'$ to the incoming amplitudes $ a_1, a_2,a_3, a_4 $. Introducing the phenomenological parameter $ \epsilon $ which measures the dephasing strength we arrive at the result \cite{twentynine}:
  \be  
\mbox{ $ s $\/ } \mbox{ = }
\left(
\begin{array}{cccc}
\mbox{\/$ 0  $} & \mbox{\/$ \sqrt{1 - \epsilon}$} &
\mbox{\/$ \sqrt{\epsilon}$} & \mbox{\/\ 0} 
\nn \\ \nn \\
 \mbox{\/$ \sqrt{1 - \epsilon}$} & \mbox{\/$ 0  $} &
\mbox{\/$ 0  $} &\mbox{\/$ \sqrt{\epsilon}$} 
\nn \\ \nn \\
\mbox{\/$ \sqrt{\epsilon}$} & \mbox{\/$ 0  $} &
\mbox{\/$ 0  $} &\mbox{\/$ - \sqrt{1 - \epsilon}$} 
\nn \\ \nn \\
\mbox{\/$ 0  $} & \mbox{\/$ \sqrt{\epsilon}$} &
\mbox{\/$- \sqrt{1-\epsilon}$} & \mbox{\/\ 0} 
\end{array}
\right)\mbox{.}
\ee  
  The parameter $ \epsilon $ takes on values within the interval $[0,1]. $ When $ \epsilon = 0 $ the bridge is detached from the scatterer, and the ET is completely coherent. In the opposite limit $ \epsilon = 1 $ the electron is surely scattered to the reservoir and then re-emitted from there, which results in complete phase randomization.

Also, the electron tunneling through a single barrier is described with the transmission and reflection amplitudes $(t $ and $ r, $ respectively). These are the matrix elements of a  $ 2 \times 2 $ matrix:
   \be 
\mbox{ $s_{el} $\/}  \mbox{ = }
\left(
\begin{array}{cc}
\mbox{\/$r $} &  \mbox{\/\ $t$ }\nonumber\\
\mbox{\/ $ t $ } &  \mbox{\/ $ r $}
\end{array}
\right) .
 \ee 
 Combining Eqs. (13) and  (14) we arrive at the expression for the scattering matrix $ S $  which gives the relation of the amplitudes $ b_1', b_2', a_3', a_4'$ to the amplitudes  $ b_1, b_2, a_3, a_4 .$ In other words, this matrix describes the ET in our system with a single-site bridge. The scattering matrix $S$ is given by \cite{twentyseven,twentynine}:
       \be 
S = Z^{-1}
\left (
\begin{array}{cccc}
r_1 + \alpha^2 r_2 & \alpha t_1 t_2 & \beta t_1 & \alpha \beta t_1 r_2
\nn \\ \nn \\
\alpha t_1 t_2 & r_2 + \alpha^2 r_1 & \alpha \beta r_1 t_2 &  \beta t_2 
\nn \\ \nn \\
\beta t_1 & \alpha \beta r_1 t_2 & \beta^2 r_1 & 
 \alpha r_1 r_2 - \alpha
\nn \\  \nn \\
\alpha \beta t_1 r_2 & \beta t_2  &  \alpha r_1 r_2 -  \alpha  & \beta^2 r_2
\nn 
\end{array}
\right )
                 \ee
  where $ Z = 1 - \alpha^2 r_1 r_2 ; \ \alpha = \sqrt {1 - \epsilon} ; \ \beta = \sqrt{\epsilon}; \ r_{1,2} $
 and $ t_{1,2} $ are the amplitude transmission and reflection coefficients for the two barriers.  
 The above expression could be generalized to include an arbitrary number of the bridge sites, as shown in  Ref. \cite{twentyseven}.  Within this approach we introduce the effective electron transmission $ T_{eff} (E) :$
  \be 
 T_{eff} \equiv \frac{J_A}{J_D}
  \ee
 where $ J_D $ is the  incoming  current from the donor and $ J_A $ is the  outgoing current from the acceptor. An electron could be injected into our system via four channels indicated in the above schematic drawing. Outgoing currents in these channels $ (J_j') $ are related to the incoming ones $ (J_i) $ by means of the transition matrix $ T\ (T_{ij} = |S_{ij}|^2 ): $
  \be 
  J_j'= \sum_i T_{ji} J_i .
  \ee
  Here, $ J_D = J_1, \ J_A = J_2'. $ To provide the charge conservation, the net current must be zero in the channels 3 and 4 connecting the system with the reservoir, so we have
  \be 
  J_3 + J_4 - J_3'- J_4'= 0.
  \ee
  There are no grounds to anticipate any difference between these two channels, therefore we assume $ J_3 = J_4. $ Also, we put $ J_2 = 0 $ presuming zero income current in the channel 2. 

Now, we use equations (17) to express outgoing currents in terms of the incoming ones, and we substitute the results into (18). This gives
  \be  
  J_3 = J_4 = J_D \frac{K_1(E)}{2 - R(E)}
   \ee
  where
  \bea 
 && K_1 (E) = T_{31} + T_{41}, \nn \\ \nn \\
 &&R(E) = T_{33} + T_{44} + T_{43} + T_{34}.
  \eea
 Inserting the expressions (19) into Eq. (17) for the current $ J_2'= J_A, $ we obtain
   \be 
 T_{eff} (E) =  T(E) + \frac{K_1 (E) K_2 (E)}{2 - R(E)}.
   \ee
  Here, the functions $ T(E),$  $ K_{2} (E) $ and $ R(E) $  introduced in the expression (21) are given by:
         \bea 
 &&  T (E) = T_{21},  \nn \\ \nn \\ 
  &&  K_2(E) = T_{23} + T_{24}.
    \eea
   To simplify further analysis we  assume both barriers to be identical, so we have $ r_1 = r_2 \equiv r $ and $ t_1 = t_2 \equiv t $. Then we easily arrive at the analytical expression for the effective transmission function:
  \be 
 T_{eff} (E) = \frac{t^2(E) (1 + \alpha^2) [1 - \alpha^2(1 - t^2(E))]}{2 [1 + \alpha^2 (1 - t^2 (E))]^2} .
  \ee

In the coherent limit $ (\epsilon = 0, \ \alpha = 1 ) $ the transmission function $ T_{eff} (E ) $ coincides with $ T(E) $ given by Eq. (11), so we can express the tunneling parameter $ \delta (E ) \equiv t^2 (E ) $ in terms of the earlier introduced Green's functions and self-energy corrections:
  \be 
 \delta (E ) = \frac{2g}{1 + g}
  \ee
  Assuming the identical barriers, the function $ g(E)$ has the form
  \be 
 g(E) = \frac{ 2\Delta}{ \sqrt{(E - E_1)^2 + \Gamma^2} }.
  \ee
 In this expression, $ \Delta_1 = \Delta_2 = \Delta, \  \Delta_{1,2} $ represent imaginary parts of the self-energy corrections introduced by Eqs. (12), and the Green's function for the single-site bridge is given by
   \be 
  G(E) = \frac{1}{E - E_1 + i \Gamma}
  \ee
 where $ \Gamma = 2\Delta  $ and $ E_1 $ is the site energy. The width of this energy level  is determined  with the parameter $ \Gamma. $
 Using these expressions, we have to keep in mind that both site energy $ E_1 $ and the parameter $ \Gamma $  change their values when we attach the bridge to the dephasing reservoir. In our system phase breaking effect originates from nuclear motions which could be described as the set of phonons  therefore $ E_1 $ and $ \Gamma $ are replaced by  \cite{thirtyfive}:
  \be 
 \tilde E_1 = E_1 + \mbox{Re} \Sigma_{ph} (E);
 \qquad \tilde
 \Gamma = \Gamma + \mbox{Im} \Sigma_{ph} (E) 
 \ee
 where $ \Sigma_{ph} (E ) $ is the electronic self-energy correction due to the electron-phonon interaction. Nevertheless, the expression (25) for the function $ g (E) $ maintains its form provided that the replacement (27) is carried out.

So, we get
   \be 
  T_{eff}(E) = \frac{g(E) (1 + \alpha^2)[g(E)(1+\alpha^2) + 1 - \alpha^2]}{[g(E)(1 - \alpha^2) + 1 + \alpha^2]^2} .
  \ee
 As it was already mentioned
 for the coherent tunneling $ (\epsilon = 0 , \ \alpha = 1) $ the transmission function $ T_{eff} (E) $ reduces to $ T(E) $. In the opposite limit of the totally incoherent transport we arrive at the expression:
  \be 
  T_{eff}(E) = \frac{g(E)}{1 + g(E)} =
\frac{2 \Delta}{\sqrt{(E - \tilde E_1 )^2 +  \Gamma^2} + 2 \Delta}.
  \ee

\begin{figure}[t] 
\begin{center}
\includegraphics[width=6.0cm,height=6.0cm]{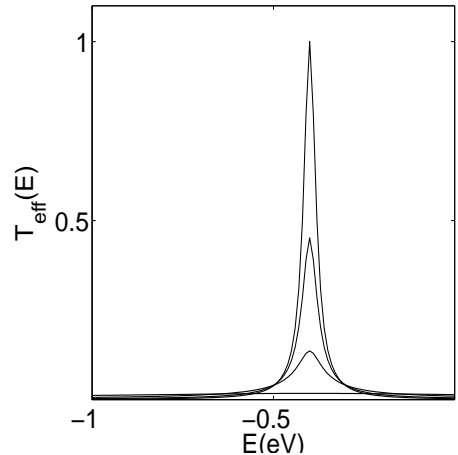}
\caption{The effective transmission function (36) versus energy. Calculation are carried out for a single electronic state bridge at $ T = 70 K, \ \tilde E_1 = - 0.4 eV, \ \Delta_1 = \Delta_2 = 10 meV.$ Curves are plotted assuming that the dephasing parameter $ \epsilon $ accepts values $ 0.0, \ 0.1 \ 0.3 $ and $ 0.7 $ (from the top the bottom).
} 
\label{rateI}
\end{center}
\end{figure}

To further simplify our analysis we proceed within the wide-band limit. Correspondingly, we assume that the self-energy terms in the expression (25)  do not depend on energy. Then the coherent electron transmission $ (\epsilon = 0) $ shows a sharp peak at the energy $ \tilde E_1 $ (see Fig. 1) which gives rise to a step-like form of the volt-ampere curve presented in the Fig. 2. When the dephasing reservoir is attached to the electronic bridge $ (\epsilon \neq 0) $ the peak in the transmission is eroded. The greater is the value of the dephasing parameter $ \epsilon $ the stronger is the erosion. When $ \epsilon $ takes on value $ 0.7 $ the peak in the electron transmission functions is completely washed out as well as the step-like shape of the $ I-V$ curve. The latter becomes linear corroborating the well-known Ohmic Law for the sequental hopping mechanism.

\begin{figure}[t] 
\begin{center}
\includegraphics[width=6.0cm,height=6cm]{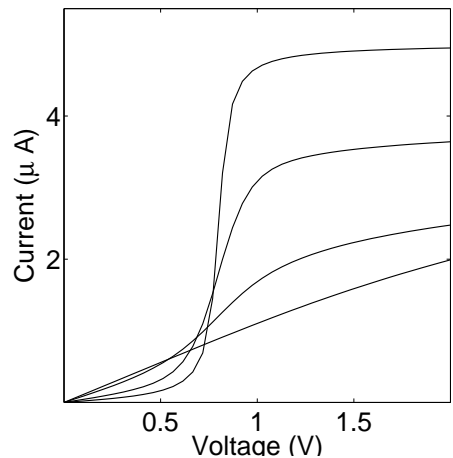}
\caption{Dephasing effect on the current-voltage characteristics. The curves are plotted for $ \epsilon = 0.0, \ 0.1, \ 0.3 $ and $ 0.7 $ from the top to the bottom. The effective electron transmission function shown in the figure 1 was used in calculations of the current.
}   
\label{rateI}
\end{center}
\end{figure}

To proceed in a more realistic way we must express the phenomenological dephasing strength $ \epsilon $ in terms of some relevant energies characterizing the ET process. To obtain the desired expression for $ \epsilon $ we compare our results with those recently presented in Ref. \cite{thirtyfive}. In that work the correction $ \delta I $ to the coherent tunnel current via a single-site bridge is calculated using 
nonequilibrium Green's functions formalism. The relevant result (See Eq. (33) in Ref. \cite{thirtyfive}) is  derived in the limit of weak electron-phonon interaction when $ \Gamma_{ph} << \Gamma \ (\Gamma_{ph} = \mbox{Im} \Sigma_{ph}). $ It is natural to assume that the dephasing strength $ \epsilon $ is small within this limit, so we expand
our expression for $ T_{eff} (E) $ in powers of $ \epsilon. $ Keeping two first terms in this expansion, we get
  \be 
  T_{eff} (E) \approx  g^2 (E) \Big (1 + \frac{\epsilon}{2} \big[1 -2 g^2 (E) \big] \Big ).
   \ee
  We employ this approximation to calculate the current through the bridge, and we arrive at the following expression for $ \delta I: $
 
 \bea 
  \delta I &=& \frac{e}{\hbar} \int\limits_{-\infty}^\infty
 dE \frac{\epsilon \Delta }{\Gamma + \Gamma_{ph}} \rho_{el} (E)
\frac{(E - \tilde E_1)^2 - 4 \Delta^2}{(E - \tilde E_1)^2 + 4 \Delta^2}
   \nn \\ \nn \\  &\times &
 \big [f(E - \mu_1) - f (E - \mu_2) \big ].
  \eea
  Here, $ \rho_{el} (E) $ is the electron density of  states at the bridge:
   \be 
 \rho_{el} (E) = - \frac{1}{\pi} \mbox{Im} G(E)
   \ee
 
Comparing the expression (31) with the corresponding result of Ref. \cite{thirtyfive} we find out that these two are consistent, and we get
  \be 
  \epsilon = \frac{\Gamma_{ph}}{\Gamma + \Gamma_{ph}}.
  \ee 
 When $ \Delta >> \Gamma_{ph} \ (\epsilon << 1)$ the bridge coupling to the dephasing reservoir is weak, and the superexchange mechanism of ET is predominating. The opposite limit $ \Delta << \Gamma_{ph}\ (\epsilon \sim 1) $ corresponds to the completely  incoherent phonon assisted ET. 
The phonon contribution to the electron linewidth depends on energy $ E. $ Starting from the general expression for $ \Gamma_{ph} (E) $  derived in Ref. \cite{twentythree}, we can write out the approximation appropriate at low temperatures within the relevant energy range $ \mu_2 < E < \mu_1, $ namely:
                 \bea 
 \Gamma_{ph}(E) &=& \pi M^2 \bigg \{ \int_0^{(\mu_1 - E)/\hbar}
d \omega \rho_{ph} (\omega ) \rho_{el} (E + \hbar \omega)
  \nn \\ \nn \\ &+&
 \int_0^{(E - \mu_2 )/\hbar}
d \omega \rho_{ph} (\omega ) \rho_{el} (E - \hbar \omega) \bigg \}.
    \eea
  Here, $ M $ is the electron-phonon coupling strength, and $ \rho_{ph} (\omega  ) $ is the phonon density of states. It includes contributions from phonons representing the thermal environment (stochastic nuclear motions) as well as from vibrational modes. 

Intensity of stochastic nuclear motions strongly depends on temperature. Therefore we may expect vibrational modes contributions to predominate in the phonon density of states at low temperatures. Assuming a single vibrational mode with the frequency $ \Omega $ to exist in the system, we have:
   \be 
\rho_{ph} (\omega) = \frac{1}{\pi\hbar} 
\frac{\gamma}{(\omega-\Omega)^2 + \gamma^2}.
  \ee
  In typical situations the linewidth $ \gamma $ is much smaller than $ \Omega $ \cite{thirtyfive}, so $ \rho_{ph} (\omega) $ exhibits rather sharp maximum at $ \omega = \Omega. $ We can roughly estimate $ \Gamma (E) $ approximating $ \rho_{ph} (\omega) $  given by Eq. (35) as $ \delta(\omega - \Omega). $ Then we obtain:\vspace{-3mm}
    \bea 
  \Gamma_{ph} (E) &\approx& \pi M^2 \big \{ \rho_{el} (E+ \hbar \Omega ) \theta (\mu_1 - \hbar \Omega - E) 
   \nn \\ 
&+& \rho_{el} (E - \hbar \Omega ) \theta (E - \mu_2 - \hbar \Omega ) \big \} 
   \eea
  where $ \theta (x) $ is the step function.

So we see that the peak in the frequency dependence of $ \rho_{ph} $ is  reflected in the energy dependencies of $ \Gamma_{ph} $ and the dephasing parameter $ \epsilon. $ Being taken into account, this energy dependence of the dephasing strength may bring some features (peaks and dips) in the first and second derivatives of the current through the bridge  $ I $ with respect to the applied voltage $ V. $ Such features were intensely studied in recent papers  (see e.g. Refs. \cite{twentythree,thirtyfive,fortyseven}), and we do not discuss the issue here. Further more, the dependence of the dephasing parameter $ \epsilon $ of energy usually does not bring significant changes in the shapes of $ I-V $ curves themselves, compared to those calculated assuming $ \epsilon $ to be a constant.
Replacing $ \rho_{el} (E \pm \hbar \Omega) $ with their maximum values we obtain that $ \Gamma_{ph} < M^2/\Delta, $ and the dephasing parameter $ \epsilon $ will take on small values provided that $ M < \Delta. $ Therefore, we have grounds to expect the dephasing rate to be reduced at low temperatures and the steps in the $ I - V $ characteristics to be distinguishable.

\section{IV. Results and Discussion}

It follows from the obtained results  that the
electron transmission function contains important information concerning intramolecular electron transfer. Under low temperatures and moderate electronic dephasing $ T_{eff}(E) $ can exhibit series of peaks. Their location is determined with contributions from donor/acceptor sites predominating the donor/acceptor coupling to the bridge at a given interval of tunneling
energy E, and with the energy spectrum of a chain of sites connecting them. Different bridge chains (pathways) contribute to the electron transmission function to different extent, but we can expect the primary pathways to predominate, and to produce the most distinguishable series of peaks in the structure of the electron transmission function in appropriate energy ranges.

The structure of $ T_{eff} (E) $ can be revealed in the low temperature current-voltage characteristics for the electronic transport through the molecular bridges. At low temperatures current-voltage curves can exhibit a step-like behavior. This originates from a step-like character of Fermi distribution functions at low temperatures along with the ''comblike'' structure of the electron transmission function $ T_{eff}(E)$. At a given voltage
$V$ the difference of Fermi functions in the Eq. (\ref{e10}) takes on nonzero values only in the interior of a certain energy range including $E_F. \!$
Within a linear approximation the size of the range $ \delta E $ is proportional to the magnitude of the voltage $V$ applied across the bridge. Therefore the magnitude of the current $I $ at a given voltage is
determined by the contributions of peaks of the electronic transmission function located in this energy range. When the applied voltage increases,
this enhances the width of the energy interval $ \delta E.$

However, the enhancement of $ \delta E $ does not immediately give rise to an enhancement of the tunneling current $ I. $ The latter abruptly changes when an extra peak of the electron transmission function appears in the interior of the energy range where $ f(E - \mu_1) - f (E - \mu_2) $ differs from zero. This explains a step-like behavior of the tunneling current as a function of the applied voltage. We also see that the steps originate from the structure of $ T_{eff}(E) $ in the foresaid energy interval. Widths of the plateaus are equal to the distances between adjacent peaks in $ T_{eff}(E) $ and magnitudes of sudden changes in the current between the
plateaus correspond to the heights and shape of these peaks.

At higher temperatures Fermi distribution functions in the integrand of the Eq. (\ref{e10}) lose their step-like character and the plateaus are washed out. Another reason for the structure of $ T_{eff}(E) $ (and $ I-V$ curves) to be eroded is the electronic phase-breaking effect which arises in complex molecules due to nuclear motions. We expect, however, that at low temperatures the electronic dephasing effects could be reduced so that the structure of $ T_{eff}(E) $ could be revealed.

On these grounds we believe that experiments on electronic transport through macromolecules at low temperatures $ (T \sim 1K )$ could be useful to obtain an additional information upon characteristics of the
intramolecular electron transfer. Namely, comparing the structure of the electron transmission function reconstructed on the basis of experimental $ I-V $ curves with that obtained as a result of calculations, we can make conclusions concerning actual primary pathways of electrons through intramolecular bridges, as well as sites of the donor/acceptor subsystems involved in the ET process at different values of the tunnel energy.
The above comparison can be succesfully performed for the molecules including bridges which can be reasonably well simulated as sets of weakly interacting chains. 

The simple model of donor-bridge-acceptor system adopted in the present work can provide us with the results suitable for a quantitative comparison with the results of proposed experiments even avoiding a proper and reliable calculation of some parameters such as the equilibrium Fermi energy of the bridge, and the effective voltage $V$ applied across the bridge.
The lack of information about proper values of $ E_F $ and $V$ produces an uncertainty in the location of the origin at the ''voltage'' axis so we cannot identify steps of the $I-V$ curves separately.
Nevertheless, changes in the values of $E_F $ and $ V $ do not influence the electron transmission function for a given molecule, therefore the structure of series of the peaks remains fixed. This enables us to identify some series of peaks analyzing sequences of widths of successive steps of the current-voltage curves. Such analysis also can give reasonable estimations for the $ E_F$ and $V$ for the chosen molecule.

In summary, the purpose of the present work is to analyze how characteristics of the low-temperature electronic transport through macromolecules can be used to get an additional information concerning
long range electron transfer. It is shown that at low temperatures and moderate electron dephasing the electron transmission function reveales a ''comblike'' series of peaks which contain information about
donor/acceptor sites effectively participating in the ET processes and primary pathways of electrons tunneling through the bridge. This important information can be obtained as a result of analysis of experimental low
temperature current-voltage characteristics for chosen macromolecules.

\section{ acknowledgments}
  
I thank G.M. Zimbovsky for help with the manuscript. This work was supported  in part by NSF Advance  program SBE-0123654 and PR Space Grant NGTS/40091.



\begin{references}


\bibitem{one} A. M. Kuznetsov and I.Ulstrup "Electron Transfer in Physics
and Biology", Wiley, England, (1999).

\bibitem{two} H.M. McConnel, J. Chem. Phys. {\bf 35}, 508 (1961).

\bibitem{three} A.L. Burin and M.A. Ratner, J. Chem. Phys. {\bf 113}, 3941 (2000). 

\bibitem{four} J.J. Storhof and C.A. Mirkin, Chem. Rev. {\bf 99}, 1849 (1999).

\bibitem{five} D.R. Bowler, J. Phys: Condens. Matter {\bf 16}, R721 (2004).

\bibitem{six} M.L. Jones, I.V. Kurnikov and D.N. Beratan, J. Phys. Chem. {\bf A106}, 2022 (2002).

\bibitem{seven} G.S.M Tong, I.V. Kurnikov and D.N. Beratan, J. Phys. Chem. {\bf B106}, 2381 (2002).

\bibitem{eight} R.A. Marcus, J. Chem. Phys. {\bf 24}, 979 (1956); {\bf
43}, 679 (1965); Annu. Rev. Phys. Chem. {\bf 15}, 155 (1964).

\bibitem{nine} N. Sutin, Acc. Chem. Res. {\bf 15}, 275
(1982); Progr. Inorg.
Chem. {\bf 30}, 441 (1983).

\bibitem{ten} R.A. Marcus and N. Sutin, Biochem. Biophys. Acta {\bf 811},
265 (1985).


\bibitem{eleven} D.N. Beratan, J.N. Betts and J.N. Onuchic, Science {\bf
252}, 128 (1991).

\bibitem{twelve} J.N. Onuchic, D.N. Beratan, J.R. Winkler and H.B. Gray,
Science {\bf 258}, 1740 (1992).

\bibitem{thirteen} C. Goldman, Phys. Rev. A {\bf 43}, 4500 (1991).


\bibitem{fourteen} S.S. Skourtis, J.J. Regan and J.N. Onuchic, J. Phys. Chem.
{\bf 98}, 3379 (1994).

\bibitem{fifteen} J.J. Regan, A.J. Di Bilio, R. Langen, L.K. Skov, J.R.
Winkler, H.B. Gray, J.N. Onuchic, Chem. Biolog. {\bf 2}, 484 (1995).

\bibitem{sixteen} J.N. Gehlen, I. Daizadeh, A.A. Stuchebrukhov and R.A.
Marcus, Inog. Chem. Acta {\bf 243}, 271 (1996).
 
\bibitem{seventeen} J. Kim and A.A. Stuchebrukhov, J. Phys. Chem. {\bf 104},
8608 (2000).

\bibitem{eighteen} Another method of transition amplitudes calculation not based on the Green's functions formalism was recently developed \cite{twenty}, and applied for studies of ET processes in protein and DNA molecules \cite{twenty,twentyone,twentytwo}.


\bibitem{nineteen} C.C. Page, C.C. Moser, X. Chen, P.L. Dutton, Nature {\bf 402}, 47 (1999). 

\bibitem{twenty} A.A. Stuchebrukhov,  J. Chem. Phys. {\bf 104}, 8424 (1996).

\bibitem{twentyone} D.M. Medvedev and A.A. Stuchebrukhov,  J. Theoretical Biology {\bf 210}, 237
(2001).

\bibitem{twentytwo} I. Daizadeh, D.M. Medvedev and A.A. Stuchebrukhov, Molecular Biology and Evolution, {\bf 19}, 406 (2002).

\bibitem{twentythree} T. Mii, S.G. Tikhodeev and H. Ueba, Phys. Rev. B {\bf 68}, 205406 (2003).

\bibitem{twentyfour} B.C. Stipe, M.A. Rezaei and W. Ho, Phys. Rev. Lett. {\bf 82}, 1724 (1999).

\bibitem{twentyfive} V.S. Panda and J.N. Onuchic, Phys. Rev. Lett. {\bf 78}, 146 (1997).

\bibitem{twentysix} V.B.P. Leite, J. Chem. Phys. {\bf 110}, 10067 (1999).

\bibitem{twentyseven} X.-Q. Li and Y.Y.Jan, J. Chem. Phys. {\bf 115}, 4169
(2001).

\bibitem{twentyeight} Y. Tanimura, V. B. P. Leite and J. N. Onuchic, J. Chem. Phys. {\bf 117}, 2172 (2002).

\bibitem{twentynine} M. Buttiker, Phys. Rev. B {\bf 33}, 3020 (1986).

\bibitem{thirty} J.L. D'Amato and H.M. Pastawski, Phys. Rev. B {\bf 41}, 7411 (1990).



\bibitem{thirtyone} V. Mujica, M. Kemp and M.A. Ratner, J. Chem. Phys. {\bf 101}, 6849 (1994).


\bibitem{thirtytwo} H. Ness and A.J. Fisher, Europhysics Letters {\bf 6}, 885 (2002); Phys. Rev. Lett. {\bf 83}, 452 (1999).

\bibitem{thirtythree} M.J. Montgomery, J. Hoekstra, T.N. Todorov and A.P. Sutton, J. Phys.: Condens. Matter {\bf 15}, 731 (2003).

\bibitem{thirtyfour} D. Segal, A. Nitzan, M.A. Ratner and W.B. Davis, J. Phys. Chem. {\bf 104}, 2790 (2000).

\bibitem{thirtyfive} See: M. Galperin, M.A. Ratner and A. Nitzan, J. Chem. Phys. {\bf 121}, 11865 (2004) and refernses therein.



\bibitem{thirtysix} For the recent review of the theory of the molecular wires conduction see: S. Datta, Nanotechnology {\bf 15}, S433 (2004).

\bibitem{thirtyseven} M.P. Samanta, W. Tian, S. Datta, J.H. Henderson and C.P. Kubiak, Phys. Rev. B {\bf 53}, R7626 (1996).

\bibitem{thirtyeight} S. Datta, W. Tian, S. Hong, R. Reinfenberger, J.H. Henderson and C.P. Kubiak, Phys. Rev. Lett {\bf 79}, 2530 (1997).


\bibitem{thirtynine} Xin--Qi Li and YiLing Yan, Appl. Phys. Lett. {\bf 79}, 2190 (2001).

\bibitem{forty} Yongqiang Xue, S. Datta, M. Ratner, J. Chem. Phys. {\bf
115}, 4292 (2001).


\bibitem{fortyone} N.D. Lang and Ph. Avouris, Phys. Rev. Lett. {\bf 84}, 358 (2000);  Phys. Rev. B {\bf 62}, 7325 (2000).

\bibitem{fortytwo} M. Di Ventra, S.T. Pantelides and N.D. Lang, Phys. Rev. Lett. {\bf 84}, 979 (2000).


\bibitem{fortythree}  N.A. Zimbovskaya and G. Gumbs, Appl. Phys. Lett., {\bf 81}, 1518 (2002).


\bibitem{fortyfour} N.A. Zimbovskaya, J. Chem. Phys.  {\bf 118}, 4 (2003).

\bibitem{fortyfive} S.S. Skourtis and S. Mukamel, J. Chem. Phys. {\bf 197}, 367 (1995).

\bibitem{fortysix} D. Segal, A. Nitzan, W.B. Davis, M.R. Wasielewski and M.A. Ratner, J. Phys. Chem, {\bf 104}, 3817 (2000).


\bibitem{fortyseven}  A. Troisi and M. Ratner, J. Chem. Phys. {\bf 118}, 6072 (2003).


\end{references}
\end{document}